%% file: main.tex
\documentclass[reprint,aps,prd,superscriptaddress,nofootinbib,floatfix,amsfonts,amssymb,amsmath]{revtex4-1}

\usepackage{graphicx,multirow,xspace,comment}
\usepackage[colorlinks=true,citecolor=blue]{hyperref}
\usepackage{breakurl}

\usepackage{xcolor}
\definecolor{red}{rgb}{1, 0, 0}

\newcommand{\nn}{\nonumber}
\newcommand{\beq}{\begin{equation}}
\newcommand{\eeq}{\end{equation}}
\newcommand{\beqa}{\begin{eqnarray}}
\newcommand{\eeqa}{\end{eqnarray}}

\newcommand{\TeV}{\rm TeV}

\def\rhobar{\bar\rho}
\def\etabar{\bar\eta}
\def\lqcd{\Lambda_{\rm QCD}}

\newcommand{\Bbar}{\,\overline{\!B}{}}
\newcommand{\Dbar}{\,\overline{\!D}{}}
\newcommand{\Kbar}{\,\overline{\!K}{}}
\def\B0bar{\Bbar{}^0}
\def\D0bar{\Dbar{}^0}
\def\K0bar{\Kbar{}^0}

\makeatletter
\g@addto@macro\bfseries{\boldmath}
\makeatother

\begin{document}

\title{New physics in \texorpdfstring{$B$}{B} meson mixing: future sensitivity and limitations}

\author{J\'er\^ome Charles$^*$}
\affiliation{CNRS, Aix Marseille Univ, Universit\'e de Toulon, CPT, Marseille, France}

\author{S\'ebastien Descotes-Genon$^*$}
\affiliation{Universit\'e Paris-Saclay, CNRS/IN2P3, IJCLab, 91405 Orsay, France}

\author{Zoltan Ligeti}
\affiliation{Ernest Orlando Lawrence Berkeley National Laboratory,
University of California, Berkeley, CA 94720}

\author{St\'ephane~Monteil$^*$}
\affiliation{ Université Clermont Auvergne, CNRS/IN2P3, LPC, F-63000 Clermont-Ferrand, France.}

\author{Michele Papucci}
\affiliation{Burke Institute for Theoretical Physics, California Institute of Technology, Pasadena, CA 91125, USA}

\author{Karim Trabelsi$^*$}
\affiliation{Universit\'e Paris-Saclay, CNRS/IN2P3, IJCLab, 91405 Orsay, France}

\author{Luiz Vale Silva$^*$}
\affiliation{IFIC, Universitat de Val\`{e}ncia - CSIC, Parc Cient\'{i}fic, Cat.\ Jos\'{e} Beltr\'{a}n 2, E-46980 Paterna, Spain\\
$^*$for the CKMfitter Group}

%\today

\begin{abstract}

The mixing of neutral mesons is sensitive to some of the highest scales probed in laboratory experiments.  In light of the planned LHCb Upgrade~II, a possible upgrade of Belle~II, and the broad interest in flavor physics in the tera-$Z$ phase of the proposed FCC-ee program, we study constraints on new physics contributions to $B_d$ and $B_s$ mixings which can be obtained in these benchmark scenarios.  We explore the limitations of this program, and identify the measurement of $|V_{cb}|$ as one of the key ingredients in which progress beyond current expectations is necessary to maximize future sensitivity.  We speculate on possible solutions to this bottleneck.  Given the current tension with the standard model (SM) in semileptonic $B$ decays, we explore how its resolution may impact the search for new physics in mixing.  Even if new physics has the same CKM and loop suppressions of flavor changing processes as the SM, the sensitivity will reach 2\,TeV, and it can be much higher if any SM suppressions are lifted.  We illustrate the discovery potential of this program.

\end{abstract}

\maketitle

\input{introduction.tex}

\input{state_of_the_art.tex}

\input{current.tex}

\input{phase1.tex}

\input{phase2.tex}

\input{phase3.tex}

\input{discussion_inputs_bottlenecks.tex}

\input{summary.tex}

%\nocite{*}
\bibliography{mixing}

\end{document}

%% file: introduction.tex
\section{Introduction}\label{sec:intro}

The mixing of neutral mesons has provided severe constraints on new degrees of freedom at high energies: since measurements of mixing and $CP$ violation in neutral kaons in the 1960s, it has provided precious information on charm and top quarks before their discovery.
The hypothesis of Kobayashi--Maskawa for the origin of $CP$ violation~\cite{Kobayashi:1973fv} observed in kaons was only tested experimentally when BaBar and Belle around 2003--2004 established $CP$ violation in good agreement with the predictions of the standard model (SM)~\cite{Charles:2004jd, Ligeti:2004ak}.  These $B$-factory results showed that the standard model (SM) source of $CP$ violation in the flavor sector was the dominant part.  However, even after BaBar and Belle, and the LHCb results of the last decade, new physics (NP) is still allowed to contribute at the 20--30\% level, compared to the SM, in flavor-changing neutral current (FCNC) processes.

Since neutral-meson mixings are FCNC processes which are suppressed in the SM, they provide strong constraints on new physics. This led to the development of numerous mechanisms to suppress such contributions, should NP exist at the TeV scale.  Low-energy supersymmetry is one example, where the ansatz of degeneracy or alignment were both motivated by constraints from neutral meson mixing and other FCNC processes.
In a large class of NP models the unitarity of the CKM matrix is maintained, and
the most significant NP effects occur in observables that vanish at tree level
in the SM~\cite{Soares:1992xi, Goto:1995hj, Silva:1996ih, Grossman:1997dd}.  In such scenarios, which encompass a large class of models, possible effects of heavy particles in each neutral meson system can be described by two real parameters,
\beq\label{param}
M_{12} = \big(M_{12}\big)_{\rm SM} \times
  \big(1 + h_{d,s}\, e^{2i\sigma_{d,s}}\big)\,,
\eeq
where $M_{12}$ relates to the time evolution of the two-state neutral meson system (for a review, see~\cite{Hocker:2006xb}).
However, the extraction of NP contribution to meson mixing is entangled with the determination of the SM parameters, namely the CKM elements.  It is not enough to measure the mixing amplitude itself, only the combination of many measurements can reveal a deviation from the SM. 
In the SM CKM fit~\cite{Hocker:2001xe, Charles:2004jd}, the constraints come from $\Delta F=1$
processes dominated by tree-level charged-current interactions, and $\Delta
F=2$ meson mixing processes, which first arise at one-loop level.  We can
modify the CKM fit to constrain new physics in $\Delta F=2$ processes,
under the assumption that it does not significantly affect the SM tree-level
charged-current interactions.  

The parameterization in Eq.~(\ref{param}) is convenient because any NP contribution to $M_{12}$ is additive, so it is easy to read off from a fit the bounds on the magnitude and the phase of the NP
contribution, or to convert the result to bounds on SMEFT operators~\cite{Jenkins:2017jig, Endo:2018gdn}.  In particular, for the NP contribution to the mixing of a
meson with $q_i\bar q_j$ flavor quantum numbers, due to the operator
\beq\label{operator}
\frac{C_{ij}^2}{\Lambda^2}\, \big( \bar q_{i,L} \gamma_\mu q_{j,L} \big)^2\,,
\eeq
where $C_{ij}$ is related to the flavour dependence and $\Lambda$ to the NP energy scale,  
one finds~\cite{Charles:2013aka} 
\begin{eqnarray}\label{hnumeric}
h &\simeq & 1.5\, \frac{|C_{ij}|^2}{|\lambda^{t}_{ij}|^2}\,
  \frac{(4\pi)^2}{G_F\Lambda^2} \simeq \frac{|C_{ij}|^2}{|\lambda^t_{ij}|^{2}} 
  \left(\frac{4.5\, \TeV}{\Lambda}\right)^2 , \nn\\ 
\sigma &=& {\rm arg}\big(C_{ij}\, \lambda_{ij}^{t*}\big) ,
\end{eqnarray}
where $\lambda^{t}_{ij} = V_{ti}^{*}\, V_{tj}$ and $V$ is the CKM matrix. 
%We used NLO expressions for the SM and LO for NP, and neglected running for NP above the weak scale.
Operators of different chiralities have conversion factors differing by ${\cal
O}(1)$ factors~\cite{Buras:2001ra}.  Minimal flavor violation (MFV), where the
NP contributions are aligned with the SM ones, correspond to
$\sigma = 0$ (mod $\pi/2$).

Substantial recent developments make it interesting to revisit the expected future sensitivities derived in 2013~\cite{Charles:2013aka}, and to explore if there are any limitations to improve the sensitivity to higher scales, from constraining NP contributions to neutral meson mixing.  The LHCb Upgrade~II has been proposed~\cite{Bediaga:2018lhg} and is likely to proceed, and discussions on a possible upgrade of Belle~II have started~\cite{belle2up}. Moreover, the FCC-ee phase of a future circular collider as a tera-$Z$ factory is generating much interest, due to
the versatility of the machine centre-of-mass energy~\cite{Blondel:2011fua}, which allows the study of all relevant electroweak thresholds ($Z$, $WW$, $ZH$, and $t \bar t$) and addresses electroweak precision physics (Higgs physics, electroweak precision observables at $Z$ pole and $WW$ thresholds) in an unrivaled way, benefiting simultaneously from both the statistics and the exquisite measurement of the beam energy at the $Z$ and $WW$ thresholds. This physics case is complemented by the unprecedented statistics attainable at the $Z$ pole (${\cal O}(5 \times 10^{12})$ $Z$ decays) that can be used for flavour physics measurements, exploiting the clean experimental environment (similar to Belle~II), and the production of all species of heavy flavors and the large boost (similar to LHCb).

This paper considers thus the following future ``Phases", as benchmarks to study:
\begin{itemize}\itemsep 0pt\setlength{\itemindent}{-.5em}
%% default indentation in revtex seems to be 1.5em
\item Phase~I: LHCb 50/fb, Belle~II 50/ab (late 2020s);
\item Phase~II: LHCb 300/fb, Belle~II 250/ab (late 2030s);
\item Phase~III: Phase~II + FCC-ee ($5 \times 10^{12}\ Z$ decays).
\end{itemize}
The ``Phase~I" benchmark here coincides with ``Stage~II" in Ref.~\cite{Charles:2013aka}, and can be seen as an update of that projection.  These data are expected to be collected by the late 2020s.  The ``Phase~II" benchmark reflects the well-developed case of the LHCb Upgrade~II~\cite{Bediaga:2018lhg} and a possible upgrade of Belle~II, which starts being discussed~\cite{belle2up}.  These data sets may be collected by the late 2030s.  Phase~III corresponds to a future circular $e^+e^-$ collider collecting $5\times 10^{12}$ $Z$ decays.  (Order $10^9 - 10^{10}$ $Z$ decays would not reach sensitivities to generic new physics in $B$ decays beyond Phase~II.)

We will focus on $B_d$ and $B_s$ mixing, and do not consider $K$ and $D$ mixing in this paper.  For $K$ mixing, it is most natural to 
parameterize NP via an additive term to the so-called $tt$ contribution to $M_{12}^{K}$ in the SM. To fully constrain its magnitude and phase two observables are needed, $\epsilon_K$ and $\Delta m_K$. However, the $tt$ contribution is only a small part of the SM contribution to $\Delta m_K$, so large reductions in lattice QCD uncertainties would be needed for meaningful improvements compared to Ref.~\cite{Charles:2013aka}.  Regarding $D$-meson mixing, the data may be accommodated by long-distance SM contributions; nevertheless important constraints on NP exist from requiring that NP contributions should not be much larger than the measurements.

In the following, Section~\ref{sec:fit} discusses the fits, starting from their inputs in Section~\ref{sec:inputs}. Section~\ref{sec:p0} discusses the current status, while Sec.~\ref{sec:p1}, \ref{sec:p2}, \ref{sec:p3} contain the results for Phases~I, II, and III, respectively. Section~\ref{sec:interpretation} discusses interpretations. 
Section~\ref{sec:bottlenecks} explores future limitations and possible ideas that may lead to improved measurements compared to current expectations.  We also explore scenarios in which NP contributes to charged current $b\to c, u$ transitions, as hinted at by the $3\sigma$ tension in measurements of the so-called $R(D)$ and $R(D^*)$ ratios of semileptonic rates~\cite{Amhis:2019ckw}.
Section~\ref{sec:sum} concludes.

%% file: state_of_the_art.tex
%\section{Generic fit for \texorpdfstring{$B_{d,s}$}{B} mixings and the state of the art}
\section{Fitting the \texorpdfstring{$B_{d,s}$}{B} mixing amplitudes %and the state of the art
}
\label{sec:fit}

\subsection{Inputs}\label{sec:inputs}

\begin{table*}
\tabcolsep 4pt
\renewcommand{\arraystretch}{1.2}
\begin{tabular}{c|cccccc}
\hline\hline
&  ~~Central~~  &  \multicolumn{4}{c}{Uncertainties}  &  ~~Reference~~ \\[-2pt] %\multirow{2}{*}{~~Reference~~}
&  values  &  Current~\cite{CKMfitterwebsite}  &  ~~Phase~I~~  &  ~~Phase~II~~  &  ~~Phase~III~~  & Phases I--III \\
\hline
$|V_{ud}|$  &  $0.97437 $ & $ \pm 0.00021$ & id & id & id & \cite{CKMfitterwebsite} \\
$|V_{us}| \, f_+^{K\to\pi}(0)$  & $0.2177 $ & $ \pm 0.0004$  & id & id & id & \cite{CKMfitterwebsite}\\
$|V_{cd}|$ & $0.2248 $ & $ \pm 0.0043$ & $\pm 0.003$ & id & id & \cite{BESIII,Ablikim:2019hff}\\
$|V_{cs}|$ & $0.9735 $ & $ \pm 0.0094$ & id & id & id & \cite{CKMfitterwebsite,BESIII,Ablikim:2019hff}\\
$\Delta m_d$ [${\rm ps}^{-1}$]   &  $0.5065 $ & $ \pm 0.0019$ & id & id & id & \cite{Amhis:2019ckw} \\
$\Delta m_s$ [${\rm ps}^{-1}$]  &  $17.757 $ & $ \pm 0.021$ & id & id & id & \cite{Amhis:2019ckw} \\
$|V_{cb}|_{\rm SL} \times 10^3$ &  \multirow{2}{*}{$42.26$}  & $ \pm 0.58 $ & $ \pm 0.60 $& $\pm 0.44$ & id & \cite{Kou:2018nap}\\
$|V_{cb}|_{W \to cb} \times 10^3$ &   & --- & --- & --- & $\pm 0.17$  & \cite{Abada:2019lih,Behnke:2013xla,FCC_Vcb}\\
$|V_{ub}|_{\rm SL} \times 10^3$ &  $3.56 $ & $ \pm 0.22$ & $ \pm 0.042$ & $ \pm 0.032 $ & id & \cite{Kou:2018nap}\\%($b\to u\ell\bar\nu$)
$|V_{ub}/V_{cb}|$ (from $\Lambda_b$) & $ 0.0842 $ & $ \pm 0.0050$ & $\pm 0.0025$ & $\pm 0.0008$ & id & \cite{Cerri:2018ypt} \\
${\cal B}(B\to\tau\nu)\times 10^4$ &  $0.83 $ & $ \pm 0.24$ & $\pm 0.04$ & $ \pm 0.02 $ & $ \pm 0.009 $ & \cite{Kou:2018nap,Abada:2019lih}\\
${\cal B}(B\to\mu\nu)\times 10^6$ & $ 0.37 $ &  ---  & $ \pm 0.03 $ & $ \pm 0.02 $ & id & \cite{Kou:2018nap}\\
\hline
$\sin 2\beta$ &  $0.680 $ & $ \pm 0.017$ & $\pm 0.005$ & $ \pm 0.002 $ & $ \pm 0.0008 $ & \cite{Cerri:2018ypt, Kou:2018nap,Abada:2019lih}\\
$\alpha\ [^\circ]$ ($ {\rm mod} \; 180^\circ $)  &  $91.9 $ & $ \pm 4.4$ & $\pm 0.6$ & id & id & \cite{Kou:2018nap}\\
$\gamma\ [^\circ]$ ($ {\rm mod} \; 180^\circ $)  &  $66.7 $ & $ \pm 5.6$ & $\pm 1$ & $ \pm 0.25 $& $ \pm 0.20 $ & \cite{Cerri:2018ypt, Kou:2018nap,Abada:2019lih}\\
$\beta_s\ [{\rm rad}]$ &  $-0.035 $ & $ \pm 0.021$ & $\pm 0.014$ & $\pm 0.004$ & $ \pm 0.002 $ & \cite{Cerri:2018ypt,Abada:2019lih}\\
$ A_{\rm SL}^d \times 10^4 $ & $-6$ & $ \pm 19 $ & $ \pm 5 $ & $ \pm 2 $ & $ \pm 0.25 $ & \cite{Amhis:2019ckw,Bediaga:2018lhg,Abada:2019lih,FCC_aSL} \\
$ A_{\rm SL}^s \times 10^5 $ & $ 3 $ & $ \pm 300 $ & $ \pm 70 $ & $ \pm 30 $ & $ \pm 2.5 $ & \cite{Amhis:2019ckw,Bediaga:2018lhg,Abada:2019lih,FCC_aSL} \\
\hline
$\bar{m}_t$ [GeV] &  $165.30 $ & $ \pm 0.32$ & id & id & $ \pm 0.020 $ & \cite{CKMfitterwebsite,Abada:2019lih} \\
$\alpha_s(m_Z)$  &  $0.1185 $ & $ %\pm 0 
\pm 0.0011$ & id & id & $ \pm 0.00003 $ & \cite{CKMfitterwebsite,Abada:2019lih} \\
$f_+^{K\to\pi}(0)$ &  $0.9681 $ & $ \pm 0.0026$ & $\pm 0.0012$ & id & id & \cite{Cerri:2018ypt}\\
$f_K$ [GeV] & $0.1552 $ & $ \pm 0.0006$ & $\pm 0.0005$ & id & id  & \cite{Cerri:2018ypt}\\
%%%$B_K$  &  $0.774 $ & $ \pm 0.012$ & $\pm 0.005$ & $\pm 0.004$ & id & \cite{Cerri:2018ypt}\\
$f_{B_s}$ [GeV] &  $0.2315 $ & $ \pm 0.0020$ & $\pm 0.0011$ & id & id & \cite{Cerri:2018ypt}\\
$B_{B_s}$ &   $1.219 $ & $ \pm 0.034$ & $\pm 0.010$ & $\pm 0.007$ & id & \cite{Cerri:2018ypt}\\
$f_{B_s}/f_{B_d}$  &  $1.204 $ & $ \pm 0.007$ & $\pm 0.005$ & id & id & \cite{Cerri:2018ypt}\\
$B_{B_s}/B_{B_d}$  &  $1.054 $ & $ \pm 0.019$ & $\pm 0.005$ & $\pm 0.003$ & id & \cite{Cerri:2018ypt}\\
$ \tilde{B}_{B_s} / \tilde{B}_{B_d} $ & $1.02 $ & $ \pm 0.05$ & $ \pm 0.013$ & id & id & \cite{Carrasco:2013zta,Bazavov:2016nty,Cerri:2018ypt} \\%$ BSts (m_b) / BStd (m_b) $
$ \tilde{B}_{B_s} $ & $0.98 $ & $ \pm 0.12$ & $ \pm 0.035$ & id & id & \cite{Carrasco:2013zta,Bazavov:2016nty,Cerri:2018ypt} \\%$ BSts (m_b) $
%\hline
$\eta_B $ & $ 0.5522 $  &  ~~$ %\pm 0 
\pm 0.0022$ ~~  & id & id & id & \cite{Buchalla:1995vs} \\
\hline\hline
\end{tabular}
\caption{Central values and uncertainties used in our analysis.
Central values have been adjusted to eliminate tensions when moving to the smaller uncertainties typical of the future projections.
The entries ``id" refer to the value in the same row in the previous column.
The assumptions entering Phase~I, Phase~II and Phase~III estimates are described in the text.
}\label{tab:CKMfitterprojinputs}\label{bigtable}
\end{table*}

We follow the CKMfitter approach for the CKM global fit~\cite{Hocker:2001xe, Charles:2004jd,
Charles:2011va,Koppenburg:2017mad} with its extension to NP in $\Delta F=2$~\cite{Charles:2004jd,Lenz:2010gu,Lenz:2012az,Charles:2013aka,Charles:2015gya} (for
other studies of such NP, see Refs.~\cite{Laplace:2002ik, Ligeti:2004ak, 
Agashe:2005hk, Ligeti:2006pm, Isidori:2010kg, Ligeti:2010ia}). We fit at the same time the CKM parameters and the NP parameters, using all the inputs available with a well-controlled sensitivity to the CKM and NP parameters.

Table~\ref{bigtable} shows all inputs and their uncertainties considered in our fits.
For an easier comparison with the present
status as of Summer 2019~\cite{CKMfitterwebsite}, the column ``Current" shows the current uncertainties (with uncertainties combined in quadrature, while in our Summer~2019 analysis statistical and theoretical uncertainties are distinguished).
We use standard SM notation for the inputs, even for  quantities which may be affected by NP in
$\Delta F=2$ processes, whose measurements have to be reinterpreted to include the NP
contributions (e.g., $\alpha$, $\beta$, $\beta_s$). Considering the difficulty to ascertain the breakdown
between statistical and systematic uncertainties in theoretical inputs for the
future projections, for simplicity, we treat all future uncertainties as
Gaussian, except for $\eta_B$ and $\alpha_s(m_Z)$ that we treat in the Rfit model~\cite{Hocker:2001xe}.

The extrapolation of lattice QCD inputs is a delicate task, since some of these results are already dominated by systematics that cannot be scaled easily over time.
Lattice QCD inputs are taken from Refs.~\cite{Kou:2018nap,Cerri:2018ypt},
with most instances in Table~\ref{bigtable} coming from the latter (in Sec.~\ref{sec:p2} we comment on the impact of using the mixing parameters in Ref.~\cite{Kou:2018nap}).
We are not aware of estimates of lattice QCD uncertainties that go farther into the future than these. The predicted lattice QCD improvements will be very important for the bag parameters related to the mixing matrix elements,
$\langle \Bbar_q | (\bar b_L \gamma_\mu q_L )^2 | B_q \rangle = (2/3)\, m_{B_q}^2 f_{B_q}^2 B_{B_q}$. 
Due to the chiral extrapolations to light quark masses, more accurate results
are available for matrix elements involving the $B_s$ meson, or for ratios
between $B_d$ and $B_s$ hadronic inputs, compared to the results for $B_d$
matrix elements.  This motivated our choice of lattice
inputs in Table~\ref{bigtable}.

The projections for the uncertainties of the exclusive semileptonic determinations of $|V_{cb}|$ and $|V_{ub}|$ combine statistical and theoretical uncertainties, the latter coming from lattice QCD extractions of the relevant form factors~\cite{Kou:2018nap}.
For Phase~I, we use the predictions labeled ``10~yr w/ EM", which should be conservative, not assuming that electromagnetic corrections are fully calculated on the lattice.  For Phase~II, we use the prediction labeled ``10~yr w/o EM", assuming that electromagnetic corrections will have been computed.
The best determinations of $|V_{cb}|$ until Phase~II come from semileptonic $B$ decays, whereas in Phase~III from $W\to \bar b c$. 
For $|V_{ub}|$, around Phase~II, its determination from $B\to \tau \bar\nu$ may become competitive with semileptonic decays.

The main uncertainties in the constraints on $\bar\rho$ and $\bar\eta$ come from the
tree-level inputs $\gamma$ and $|V_{ub}/V_{cb}|$. The combination of measurements
$\gamma(\alpha)=\pi-\beta-\alpha$, which is not affected by NP in $\Delta
F=2$~\cite{Soares:1992xi}, is significant in the current average of the $\gamma$ constraint, but it diminishes in importance at Phase~I and especially beyond that   (the determination of $\alpha$
from $B\to \rho\rho,\ \rho\pi,\ \pi\pi$ is only affected by NP in electroweak penguins~\cite{Charles:2017evz}).
The improvements in $\gamma$ beyond Phase~I, shown in Table~\ref{bigtable}, assume the so-called model-independent measurement, without charm factory input~\cite{Anton}.
The fits include the constraints from the measurements of $A_{\rm
SL}^{d,s}$~\cite{Laplace:2002ik, Lenz:2010gu}, but not their linear
combination~\cite{Abazov:2011yk} nor $\Delta\Gamma_s$, whose effects on
the future constraints on NP studied in this paper are small.

Initial investigations at the physics case of FCC-ee are gathered in Ref.~\cite{Abada:2019lih}, and provide the starting ground of the present study. The inputs in Table~\ref{bigtable} correspond to the actual sensitivity studies performed so far, which are only a subset of the observables to be improved.
Most inputs considered in this work for Phase~III are obtained from extrapolations (scaling to luminosity) of the current precision or sensitivity of the measurements obtained at Belle~II and LHCb~\cite{Abada:2019lih}.
Some comments are, however, in order for two of them: $|V_{cb}|$ and $A_{\rm SL}$. The $|V_{cb}|$ sensitivity is derived from the counting of the $W$ decays selected with two jets satisfying $b$-tagging and $c$-tagging algorithms, which performance is given in Refs.~\cite{Behnke:2013xla,FCC_Vcb}. It is already observed from this state-of-the-art starting point that the precision on the $|V_{cb}|$ matrix element is improved by a factor 3--4. The precision of the semileptonic $CP$ asymmetries are obtained from a fast simulation study~\cite{FCC_aSL}. A similar method as employed by LHCb~\cite{Aaij:2016yze} is considered, using a squared-cut based selection of the decays $B_s \to D_s \ell \nu X$, but enhanced to decays of $D_s$ containing $\pi^0$ and $K_S$.   The obtained statistical precision is a few times $10^{-5}$, which makes possible to attain the SM value. However, the detection asymmetries are expected to
be a limitation of the method, at a level comparable to the statistical uncertainty.

%% file: current.tex
\subsection{\boldmath Current status}
\label{sec:p0}

\begin{figure}[tb]
\includegraphics[width=.95\columnwidth,clip,bb=15 15 550 470]{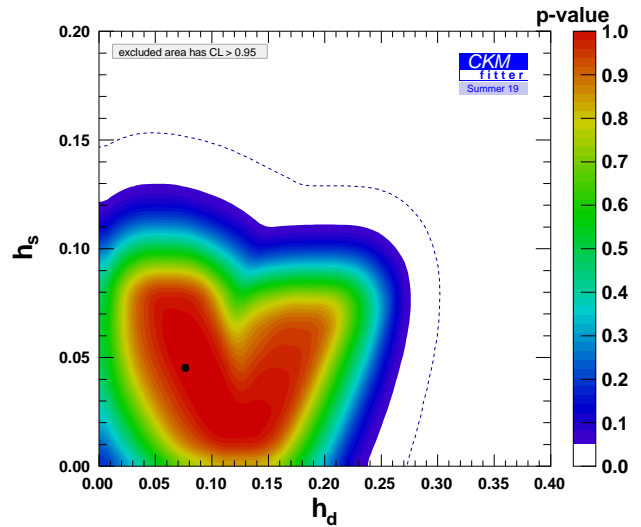}
\caption{Current sensitivities to $h_d-h_s$ in $B_{d}$ and $B_{s}$ mixings as of Summer 2019 \cite{CKMfitterwebsite}. The black dot indicates the best-fit point, and
the dotted curve shows the 99.7\%\,CL ($3\sigma$) contour.
}
\label{hdhs_0}
\end{figure}

The present constraints on the magnitudes of NP contributions to the $B_d$ and $B_s$ mixing amplitudes are shown in Fig.~\ref{hdhs_0}, with inputs corresponding to the Summer 2019 version of the CKMfitter Collaboration updates~\cite{CKMfitterwebsite}, to which we add the inputs $ A_{\rm SL}^d = 0.0000 \pm 0.0019 $ and $ A_{\rm SL}^s = +0.0016 \pm 0.0030 $ (with +6.6\% correlation)~\cite{Amhis:2019ckw}. In the SM fit ($h_d = h_s =0$) the pulls of the observables $ \Delta m_d $ and $ \Delta m_s $ are 1.7 and 1.3~$\sigma$, respectively.  Allowing for NP contributions, the fit shown in Fig.~\ref{hdhs_0} favors $h_d$ and $h_s$ somewhat away from the origin, alleviating the pulls of $ \Delta m_d $ and $ \Delta m_s $ to 0.4 and 0.2~$\sigma$, respectively. Fig.~\ref{hdhs_0} shows agreement with the SM hypothesis at $ \sim 1 \sigma$.

In the NP scenario, the 1$\sigma$ intervals for the Wolfenstein parameters are
\begin{eqnarray}
&& A = 0.813^{+0.016}_{-0.015} \,, \qquad \lambda = 0.224835^{+0.000255}_{-0.000059} \,, \nonumber\\
&& \bar{\rho} = 0.122^{+0.025}_{-0.022} \,, \qquad \bar{\eta} = 0.371^{+0.022}_{-0.015} \,.
\end{eqnarray}
Note that the uncertainties of $\bar{\rho}$ and $\bar{\eta}$ increase by about a factor of 3 compared to the fits assuming the SM, while for the NP parameters we obtain
\begin{eqnarray}
&& h_d = 0.075^{+0.153}_{-0.064} \,, \qquad h_s = 0.048^{+0.048}_{-0.048} \,, \nonumber\\
&& \sigma_d = -1.40^{+0.97}_{-0.23} \,,
\end{eqnarray}
with $\sigma_s$ unconstrained at $1\sigma$.
The plot in Fig.~\ref{hdhs_0}
is obtained by treating
$\rhobar$, $\etabar$, and the other physics parameters not shown as nuisance parameters.
This corresponds to the case of generic NP, ignoring possible
model-dependent relations between different $\Delta F=2$ transitions.
The constraint from $\epsilon_K$ has negligible impact throughout this paper when no NP in the kaon sector is considered; when NP in this sector is allowed as mentioned in the Introduction, $\epsilon_K$ probes NP mediating $\Delta S=2$ transitions, with no impact whatsoever on our analyses.
One can see from Fig.~\ref{hdhs_0} that LHCb measurements
have imposed comparable constraints on NP in $B_{s}$ mixing to those in the
$B_{d}$ system.  This qualitative picture will continue to hold in the future.

To estimate and plot future sensitivities for our Phase~I, II, and III benchmarks, we adjusted the central values of the input measurements to their best fit values in the SM global fit of 2019, in order to eliminate tensions when moving to smaller uncertainties in the future scenarios. The effect of adjusting the central values is illustrated by the top left plot in Fig.~\ref{hdhs}, which shows the fit with the adjusted central values of Table~\ref{bigtable} and the same uncertainties as in Fig.~\ref{hdhs_0}. By construction, the $p$-value in Fig.~\ref{hdhs} is maximal at $h_d = h_s =0$. It turns out that both fits yield similar $3 \sigma$ bounds on $h_d$ and $h_s$.

%% file: phase1.tex
\subsection{Phase I exploration}
\label{sec:p1}

\begin{figure*}[t]
\includegraphics[width=.95\columnwidth,clip,bb=15 15 550 470]{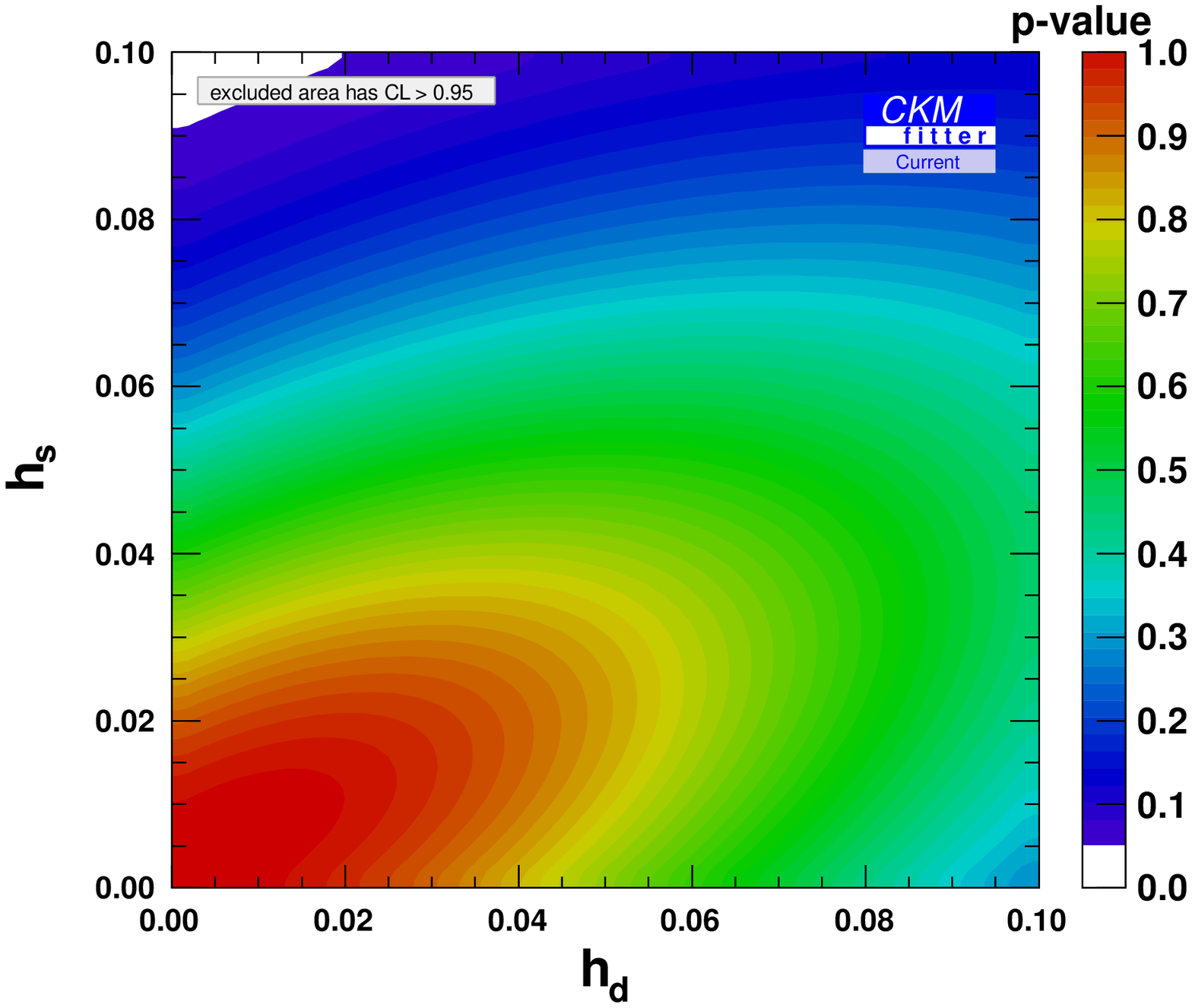}
\hfil
\includegraphics[width=.95\columnwidth,clip,bb=15 15 550 470]{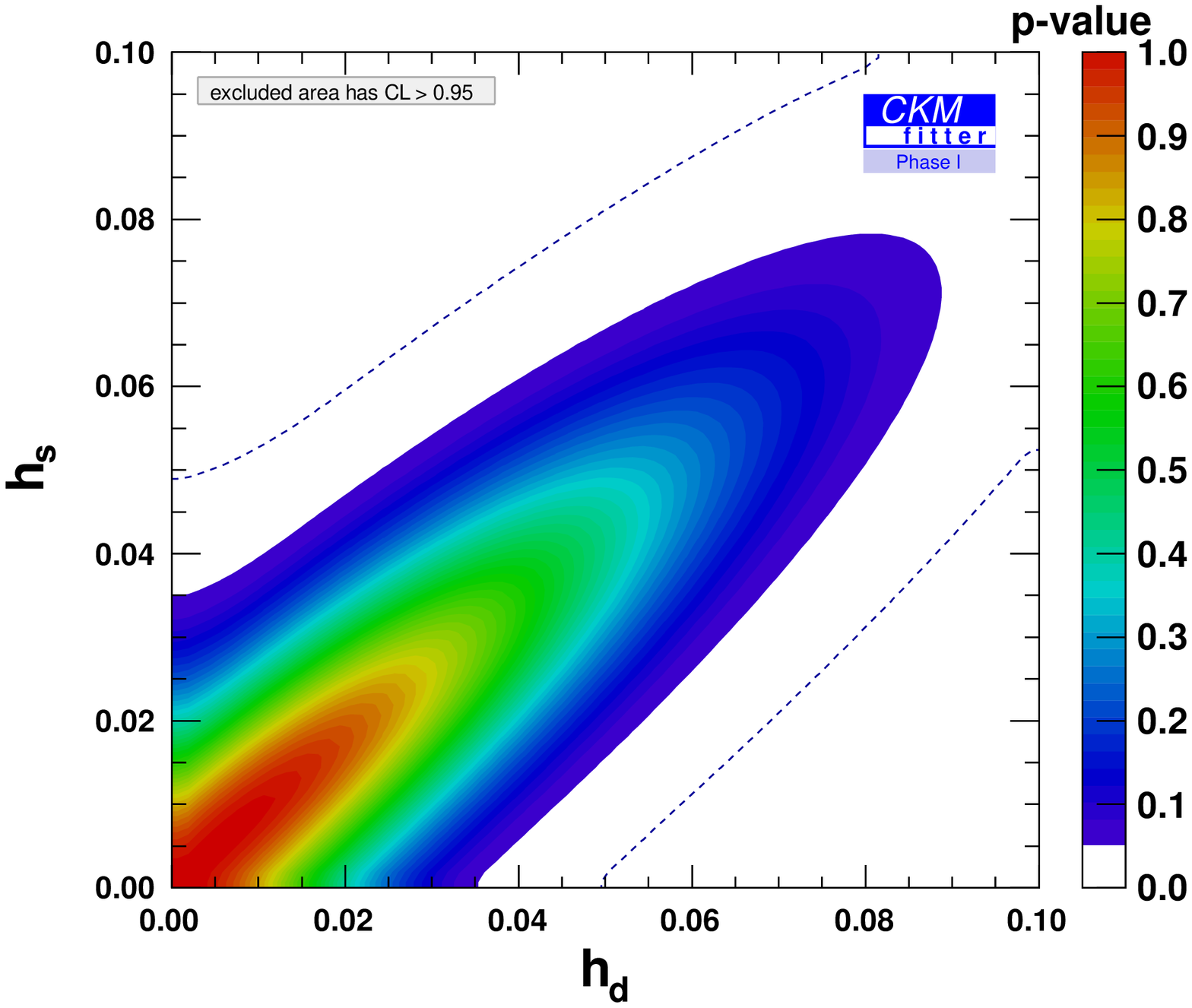}
\\
\includegraphics[width=.95\columnwidth,clip,bb=15 15 550 470]{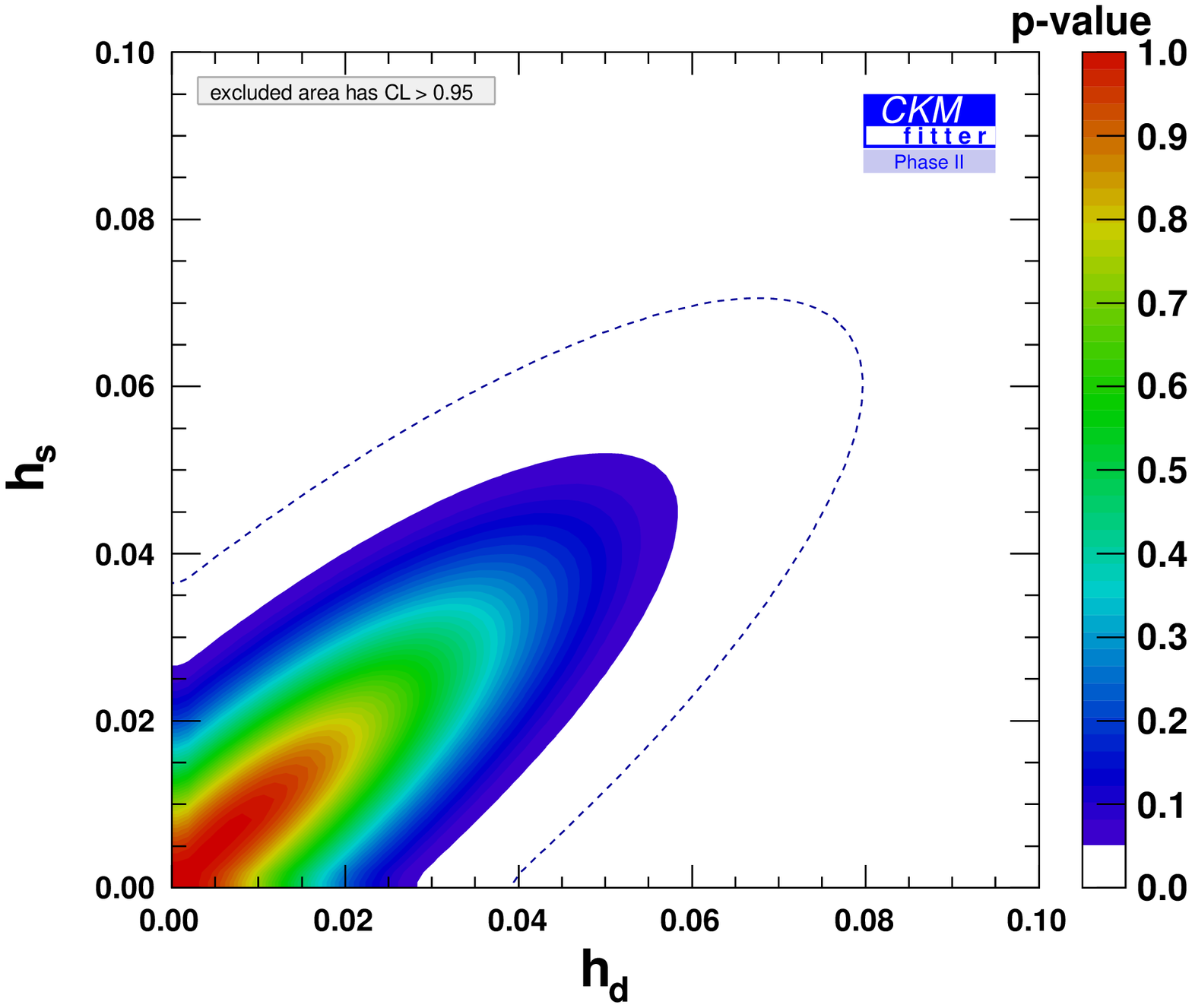}
\hfil
\includegraphics[width=.95\columnwidth,clip,bb=15 15 550 470]{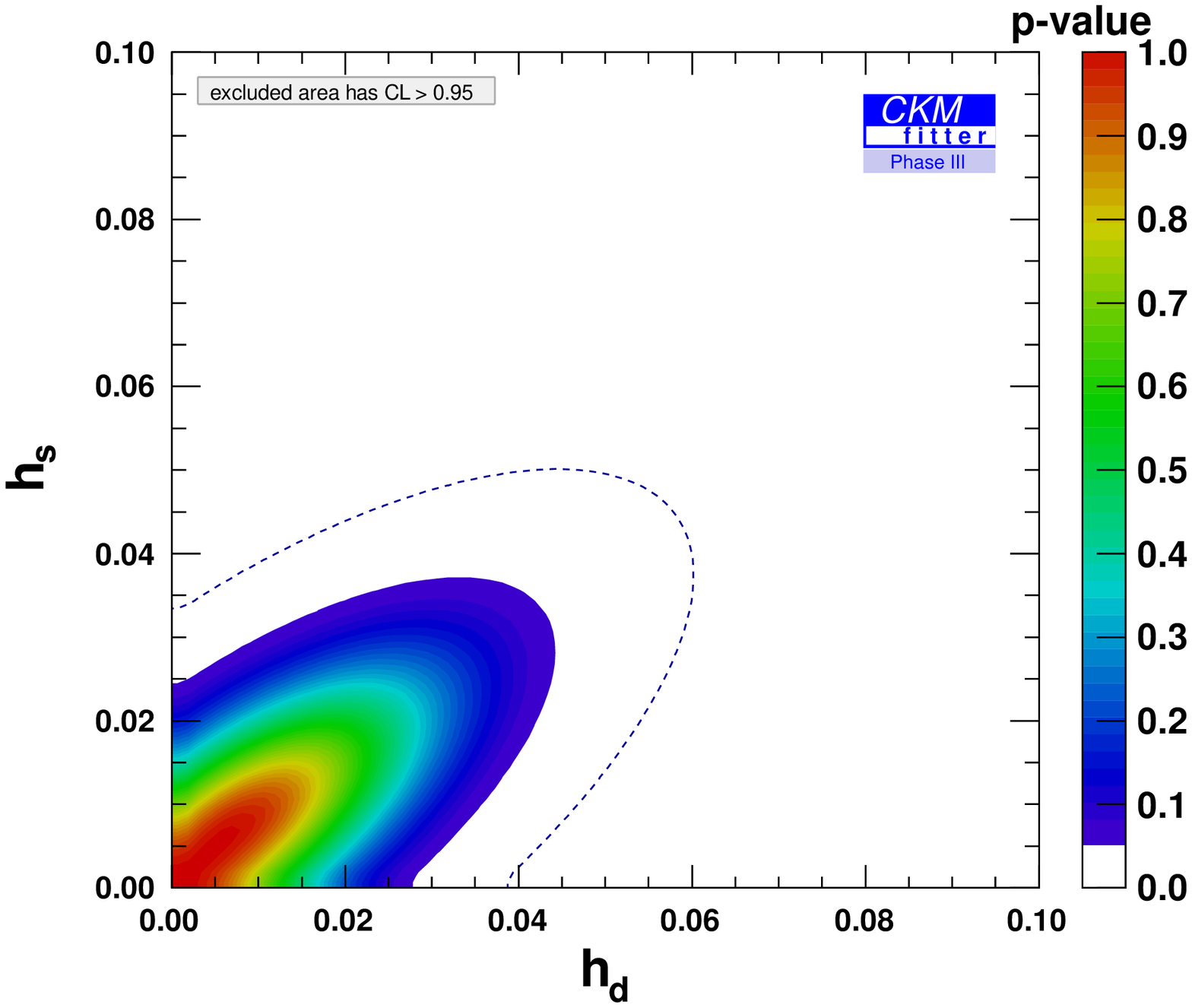}
\caption{Current (top left), Phase~I (top right), Phase~II (bottom left), and Phase~III (bottom right) sensitivities to
$h_d-h_s$ in $B_{d}$ and $B_{s}$ mixings, resulting from the data shown in Table~\ref{bigtable} (where central values for the different inputs have been adjusted).
The dotted curves show the 99.7\%\,CL ($3\sigma$) contours.
}
\label{hdhs}
\end{figure*}

\begin{table*}
\tabcolsep 4pt
\renewcommand{\arraystretch}{1.2}
\begin{tabular}{c|c|c}
\hline\hline
%	\cline{2-3}
\multirow{2}{*}{Fit description (Phase I)} & \multicolumn{2}{c}{Sensitivities at $1\sigma$} \\
  & $ h_d $ & $ h_s $ \\
	\hline
	main fit & $ [0, 0.040] $ (100\%) & $ [0, 0.036] $ (100\%) \\
	\hline
	no $ \{ f_{B_s} , f_{B_s} / f_{B_d} , B_{B_s} , B_{B_s} / B_{B_d} \} $ uncertainties & $ [0, 0.036] $ (90\%) & $ [0, 0.033] $ (92\%) \\
	\hline
	no $ \eta_B $ uncertainty & $ [0, 0.035] $ (88\%) & $ [0, 0.031] $ (86\%) \\
	\hline
	no $ \{ f_{B_s} , f_{B_s} / f_{B_d} , B_{B_s} , B_{B_s} / B_{B_d} , \eta_B \} $ uncertainties & $ [0, 0.032] $ (80\%) & $ [0, 0.029] $ (81\%) \\
	\hline\hline
\end{tabular}
\caption{The role of input uncertainties in the Phase~I results, for LHCb with 50/fb and Belle~II with 50/ab. The displayed $h_{d,s}$ ranges are at $1\sigma$, and percentages correspond to the relative uncertainty with respect to the main fit.}
\label{tab:PhaseI_d_sector}
\end{table*}

\begin{table*}[tbh]
\tabcolsep 4pt
\renewcommand{\arraystretch}{1.2}
\begin{tabular}{c|c|c|c}
\hline\hline
\multirow{2}{*}{Fit description (Phase II)}  & \multicolumn{2}{c|}{Sensitivities at $1\sigma$} & \multirow{2}{*}{~Plot in Fig.~\ref{hdhs_small_uncs}~} \\
  & $ h_d $ & $ h_s $ & \\
	\hline
	main fit & $ [0, 0.028] $ (100\%) & $ [0, 0.025] $ (100\%) & top left \\
	\hline
	no $ \{ f_{B_s} , f_{B_s} / f_{B_d} , B_{B_s} , B_{B_s} / B_{B_d} \} $ uncertainties & $ [0, 0.024] $ (86\%) & $ [0, 0.023] $ (92\%) & --- \\
	\hline
	no $ \eta_B $ uncertainty & $ [0, 0.024] $ (86\%) & $ [0, 0.021] $ (84\%) & --- \\
	\hline
	no $ \{ f_{B_s} , f_{B_s} / f_{B_d} , B_{B_s} , B_{B_s} / B_{B_d} , \eta_B \} $ uncertainties & $ [0, 0.020] $ (71\%) & $ [0, 0.019] $ (76\%) &  top right \\
	\hline
	$ \delta_{tot} | V_{cb} |_{\rm SL} / 20 $ & $ [0, 0.022] $ (79\%) & $ [0, 0.018] $ (72\%) &  bottom left \\
	\hline
	$ \delta_{tot} | V_{cb} |_{\rm SL} / 20 $, \& ``no theor.\ uncert." & $ [0, 0.0096] $ (34\%) & $ [0, 0.0061] $ (24\%) & bottom right \\
	\hline
	$ \{ \delta_{tot} | V_{ub} |_{\rm SL}, \delta_{tot} | V_{cb} |_{\rm SL}, \delta \sin (2 \beta), \delta \sin (2 \gamma) \} / 20 $, \& ``no theor.\ uncert." & $ [0, 0.0089] $ (32\%) & $ [0, 0.0061] $ (24\%) & --- \\
\hline\hline
\end{tabular}
	\caption{The role of input uncertainties in Phase~II results, for LHCb with 300/fb and Belle~II with 250/ab. We analyze the impact on bounds for $h_d$ and $h_s$ when:
	a) we 
	reduce by a factor of 20 the uncertainty of various key quantities for calculating $\Delta m_d$ and $ \Delta m_s$
	and b)
	the $\{ f_{B_s},\ f_{B_s} / f_{B_d} ,\ B_{B_s} ,\ B_{B_s} / B_{B_d} ,\ \eta_B \} $ uncertainties are set to zero (also denoted as ``no theor.\ uncert."). Percentages correspond to the relative uncertainty with respect to the main fit.}
	\label{tab:PhaseII_reducing_uncs}
\end{table*}

As indicated in Table~\ref{tab:CKMfitterprojinputs}, compared to the current status, the uncertainties of many nonperturbative theoretical inputs are anticipated to be improved by a factor of at least 1.5, up to 4. In particular, uncertainties of the bag parameters and decay constants, necessary for predicting the mass differences of the two $B_d$ and $B_s$ mass eigenstates, will all go below the percent level. At the same time, Belle~II will improve the determinations of the CKM matrix elements $|V_{cb}|$ and $|V_{ub}|$, by measuring the semileptonic channels $B \to D^{(\ast)} \ell \bar\nu$, and $B \to \pi \ell \bar\nu$. The LHCb collaboration has measured $|V_{cb}|$ for the first time at a hadronic machine~\cite{Aaij:2020hsi} and is expected to contribute to the final precision of the world average. Yet, this is not taken into account in the anticipated precision of this observable considered here. Moreover, the uncertainties in the determinations of the angles of the $B_d$ unitary triangle will reach around the $1^\circ$ level.

These improvements on theoretical inputs and data translate into much better constraints on the $h_d - h_s$ plane parameterizing the size of NP in $B_s$ and $B_d$ meson-mixing, as seen from the top right plot in Fig.~\ref{hdhs}, which assumes that future measurements remain consistent with the SM. These results are similar to the ``Stage~II" scenario shown in Ref.~\cite{Charles:2013aka}, which corresponded to the same projected LHCb and Belle~II integrated luminosities.

Table~\ref{tab:PhaseI_d_sector} illustrates the effects of reducing the uncertainties of the nonperturbative and perturbative theoretical inputs involved in the predictions of the mass differences $\Delta m_d$ and $\Delta m_s$, where we explored the consequences of eliminating their uncertainties. This table shows that setting to zero the uncertainties of the nonperturbative or the perturbative theoretical inputs have similar impacts on the allowed ranges of the NP parameters $h_d$ and $h_s$, with an improvement of about 10\% for~each.

%% file: phase2.tex
\subsection{Phase II exploration}
\label{sec:p2}

\begin{figure*}[t]
\includegraphics[width=.95\columnwidth,clip,bb=15 15 550 470]{hdhs_PhaseII_zoom}
\hfil
\includegraphics[width=.95\columnwidth,clip,bb=15 15 550 470]{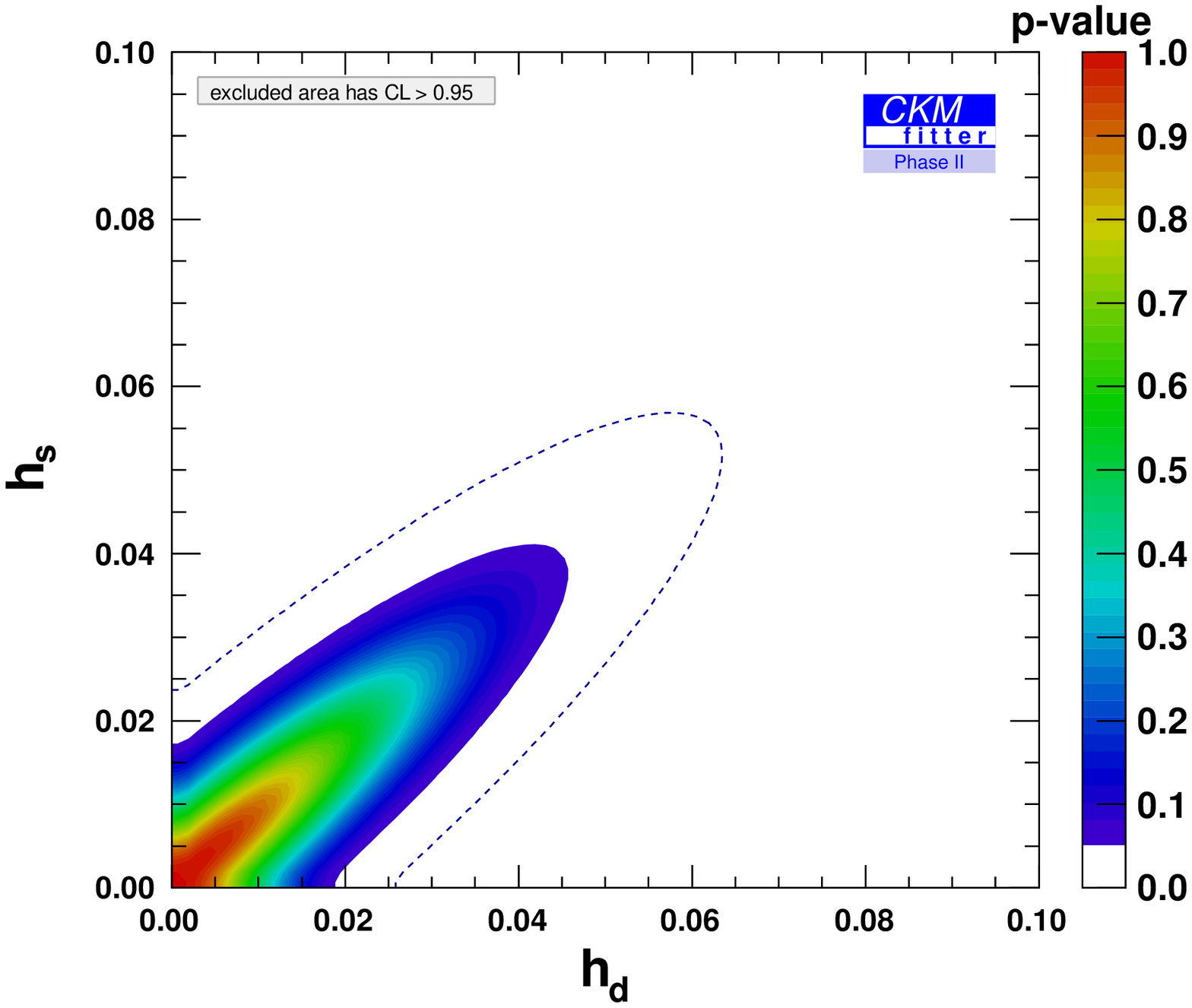}
\\
\includegraphics[width=.95\columnwidth,clip,bb=15 15 550 470]{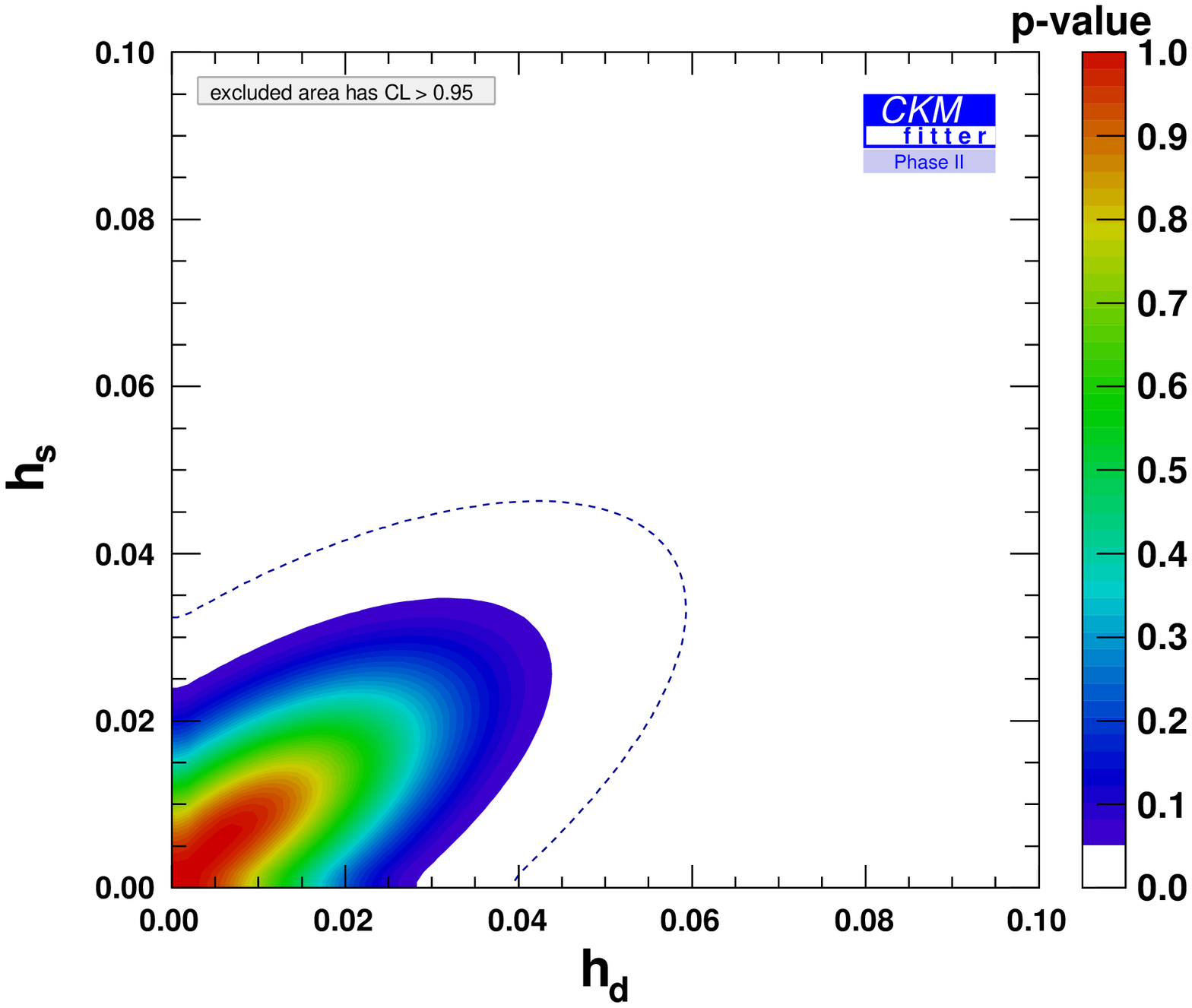}
\hfil
\includegraphics[width=.95\columnwidth,clip,bb=15 15 550 470]{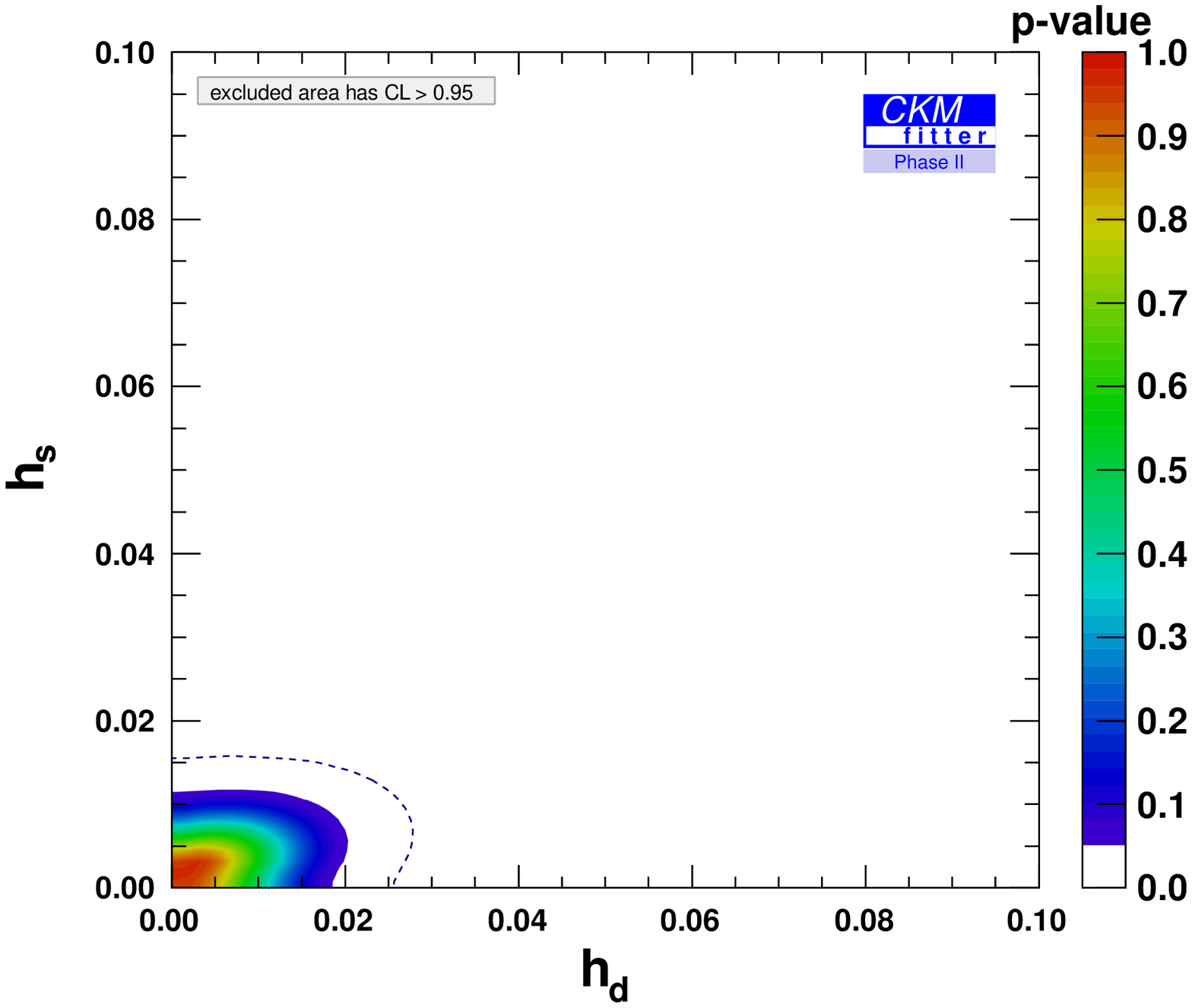}
\caption{Impact of improving key uncertainties in Phase~II: nominal Phase~II plot (top left, same as bottom left in Fig.~\ref{hdhs}), case in which uncertainties from the perturbative and nonperturbative theoretical inputs $\{ f_{B_s},\ f_{B_s} / f_{B_d} ,\ B_{B_s} ,\ B_{B_s} / B_{B_d} ,\ \eta_B \} $ are eliminated (top right), case in which the extraction of $ \delta_{tot} | V_{cb} |_{\rm SL} $ is improved by a factor 20 (bottom left, note the similarity with Phase~III in Fig.~\ref{hdhs}), combining improvements on theoretical inputs and $ \delta_{tot} | V_{cb} |_{\rm SL} $ (bottom right).
}
\label{hdhs_small_uncs}
\end{figure*}

We now shift to the sensitivity to NP achievable in Phase~II, shown in the bottom left plot in Fig.~\ref{hdhs}. As seen from Table~\ref{tab:CKMfitterprojinputs}, some key quantities such as $\phi_s$, $\gamma$ or $\beta$ will be much more precisely determined (typically by a factor of 3). Yet some other key quantities will only be slightly improved.  This is the case for the bag parameters, and also for the uncertainties in the extractions of $|V_{cb}|$ and $|V_{ub}|$ from semileptonic decays.  The reliability of these extrapolations (pertinent for the late 2030s) is necessarily less good than for Phase~I (i.e., late 2020s).

As shown in Fig.~\ref{hdhs}, the constraints on $h_d$ and $h_s$ will improve again  between Phase~I and Phase~II, although the improvement is smaller than that between the Summer~2019 situation and Phase~I. This is caused only in part by the slight pull away from the SM seen in the Summer~2019 fit in Fig.~\ref{hdhs_0}.

To understand the future limitations, we compare in Table~\ref{tab:PhaseII_reducing_uncs} the impact on the sensitivity to $h_{d,s}$ when improving or eliminating the uncertainties of some key quantities in the computations of $\Delta m_d$ and $\Delta m_s$.  As this table shows, $|V_{cb}|$ plays a central role: neglecting its uncertainty (implemented by a factor of 20 reduction), sensitivities improve by about 25\%, and the improvements increase up to 70\% when eliminating simultaneously the uncertainties coming from perturbative and nonperturbative theoretical inputs for meson mixing, as illustrated in Fig~\ref{hdhs_small_uncs}. Recall that a precise determination of $|V_{cb}|$ amounts to a precise determination of the Wolfenstein parameter $A$. Therefore, our findings imply that the remaining Wolfenstein parameters $\bar\rho$ and $\bar\eta$ are precise enough at Phase~II, while the determination of $A$ is a bottleneck for constraining further $h_d$ and $h_s$. The effect of improving only the theoretical inputs is also shown in Table~\ref{tab:PhaseII_reducing_uncs}, where an improvement in sensitivity by roughly 25\% is seen when eliminating the uncertainties of all theoretical inputs for meson mixing.

We should emphasize that many other measurements will improve substantially in Phase~II, e.g., the uncertainties of ${\cal B}(B_s\to\mu\mu)$ and ${\cal B}(B_d\to\mu\mu)$ are expected to reach 5\% and 10\%, respectively~\cite{Cerri:2018ypt}.  They will also provide high sensitivity to (other types of) NP, but do not impact the constraints on $h_{d,s}$ and $\sigma_{d,s}$.

Note also the importance of lattice QCD uncertainties.  Their predicted improvements are more uncertain the more one extrapolates into the future.  Two sets of recent predictions for the decay constants and bag parameters are shown in Table~\ref{tab:lqcdfuture}.  Our results in Fig.~\ref{hdhs} are based on Ref.~\cite{Cerri:2018ypt} for these inputs. Using instead the ``10yrs w/o EM" values from Ref.~\cite{Kou:2018nap} for Phase~II (assuming that electromagnetic effects will be included on the Phase~II timescale) would yield very similar results, since form factor projections are more optimistic, while the opposite holds for the bag parameters. The combination of Refs.~\cite{Cerri:2018ypt,Kou:2018nap} (i.e., considering the most optimistic projections in Table~\ref{tab:lqcdfuture}) leads to a slight improvement in the sensitivities to $h_d$ and $h_s$, and a strong improvement in their correlation, due to the significantly smaller uncertainty of $f_{B_s}/f_{B_d}$ with respect to \cite{Cerri:2018ypt}.

\begin{table}[b]
\tabcolsep 4pt
\begin{tabular}{c|cc|cc}
\hline\hline
\multirow{3}{*}{Uncertainties}  &  \multicolumn{2}{c|}{Ref~\cite{Cerri:2018ypt} (LHCb)} &  \multicolumn{2}{c}{Ref.~\cite{Kou:2018nap} (Belle~II 10 yrs)} \\
&  2025  &  2035  &  \multirow{2}{*}{``w/ EM"}  &  \multirow{2}{*}{``w/o EM"} \\  &  (23/fb)  &  (300/fb)~  &  \\
\hline
$\delta f_{B_s}$\,[GeV]  &  0.0011  &  0.0011  &  0.0024  &  0.00074 \\
$\delta (f_{B_s}/f_{B_d})$  &  0.005  &  0.005  &  0.012  &  0.0014  \\
$\delta B_{B_s}$  &  0.010  &   0.007  &  0.018  &  0.012 \\
$\delta (B_{B_s}/B_{B_d})$  &  0.005  &  0.003  &  0.013  &  0.0072 \\
\hline\hline
\end{tabular}
\caption{Predictions for future lattice QCD uncertainties.}
\label{tab:lqcdfuture}
\end{table}

%% file: phase3.tex
\subsection{Phase III exploration}
\label{sec:p3}

The sensitivity achievable in Phase~III is displayed in the bottom right plot in Fig.~\ref{hdhs}.  No improvement in lattice QCD uncertainties is used in going from Phase~II to Phase~III, since we are not aware of any predictions for the relevant time frame. Hence, these sensitivity projections are probably (very) conservative. The observed improvement in sensitivity to $h_d$ and $h_s$ from Phase~II to Phase~III is therefore solely related to the improvement in $|V_{cb}|$ precision at FCC-ee. The projections provided at Phase~II and even more at Phase~III support the need for simultaneously improving the CKM normalisation and the hadronic parameters describing the mixing, to fully exploit the precision of the CKM observables at these time frames. Any improvements beyond what can currently be anticipated would make the Phase~III sensitivity better than plotted in Fig.~\ref{hdhs}.
It should also be emphasized that the FCC-ee program has a much broader scope than the study discussed in this paper.

\section{Interpretations}
\label{sec:interpretation}

\begin{table}[bth]
\begin{tabular}{c|cccc}
\hline\hline
Sensitivities~  &  ~Summer~2019~  &  ~Phase~I~  &  ~Phase~II~  &  ~Phase~III \\
\hline
$h_d$  &  0.26  &  0.073  &  0.049  &  0.038  \\
$h_s$  &  0.12  &  0.065  &  0.044  &  0.031  \\
\hline\hline
\end{tabular}
\vspace*{-6pt}
\caption{Current and future 95\%~CL sensitivities to $h_{d,s}$, assuming unrelated NP contributions in $B_{d,s}$ mixings.
}
\label{tab:hds95}
\end{table}

\begin{table*}[tbh]
\tabcolsep 4pt
\begin{tabular}{c|c||c|c|c|c|c|c}
\hline\hline
\multirow{2}{*}{Couplings}  &  ~NP loop~  &  
  \multicolumn{2}{c|}{~Sensitivity for Summer~2019 [TeV]~} &  \multicolumn{2}{c|}{~Phase~I Sensitivity [TeV]~} &
  \multicolumn{2}{c}{~Phase~II Sensitivity [TeV]~} \\
&  order  &  ~$B_d$ mixing~  &  $B_s$ mixing   &  ~$B_d$ mixing~  &  $B_s$ mixing  &  ~$B_d$ mixing~  &  $B_s$ mixing \\
\hline
$|C_{ij}| = |V_{ti}V_{tj}^*|$ & tree level 
  & 9 & 13 & 17 & 18 & 20 & 21 \\
(CKM-like)  &  one loop  
  & 0.7 & 1.0 & 1.3 & 1.4 & 1.6 & 1.7 \\
\hline
$|C_{ij}| = 1$  &  tree level 
  & $1\times 10^3$ & $3\times 10^2$ & $2\times 10^3$ & $4\times 10^2$ & $2\times 10^3$ & $5\times 10^2$ \\
(no hierarchy)  &  one loop 
  & 80 & 20 & $2\times 10^2$ & 30 & $2\times 10^2$ & 40 \\
\hline\hline
\end{tabular}
\caption{The scale of the operator in Eq.~(\ref{operator}) probed (in TeV, at 95\%~CL) by $B_d$ and $B_s$ mixings at present, at Phase~I, and Phase~II, if the NP contributions in the two meson mixings are unrelated. The impact of SM-like hierarchy of couplings and/or loop suppression is shown.
}
\label{tab:scales}
\end{table*}

The 95\%~CL sensitivities to $h_d$ and $h_s$ obtained above are summarized in Table~\ref{tab:hds95}.
The energy scales probed by meson mixing can be doubled due to the anticipated improvement in the sensitivity to $h_d$, going from
the current constraints to Phase~I (improvement by more than a factor of 3) and to Phase~II (factor of 5).  These improvements compare well with those anticipated in the NP reach of the HL-LHC, during the same time frame.

The sensitivities to $h_{d,s}$ are straightforward to convert to the scales of BSM operators probed.  While BSM models may generate (combinations of) several distinct dimension-6 four-fermion operators contributing to $B-\Bbar$ mixing, for illustration we only calculate here the sensitivities to the scales of the operators which occur in the SM, shown in Eq.~(\ref{operator}).  We use Eq.~(\ref{hnumeric}) and distinguish several scenarios.  For NP with flavor structure independent of the SM Yukawa couplings, we set $|C_{ij}|$ to unity.  Many NP models contain suppressions of flavor-changing processes similar to the SM, in which case $|C_{ij}| = |\lambda^t_{ij}|$ may be appropriate (we use Ref.~\cite{CKMfitterwebsite} for the numerical values of $|\lambda_{ij}^t|$).  For NP contributions that occur at tree level, the $(4\pi)^2$ factor in Eq.~(\ref{hnumeric}) is present, while it should be removed if the NP contribution is generated at the one-loop level (similar to the SM box diagrams).  

The resulting sensitivities to NP energy scales are shown in Table~\ref{tab:scales} up to Phase~II. The scales probed at Phase~III are not shown, since they are primarily dependent on not yet estimated lattice QCD improvements at this time frame.
Nevertheless, we note that FCC-ee precision measurements would improve significantly the mixing analyses if the bottlenecks that we identified ($|V_{cb}|$ and lattice QCD parameters) can be addressed.

One sees that even if NP contains the same CKM suppressions of $\Delta F=2$ transitions as those present for the SM contributions, as well as a one-loop suppression, both of which occur for many NP models which are in the LHC energy range, the scales probed by the mixing constraints are still at the 1--2~TeV range.  These are comparable to gluino masses explored at the HL-LHC, and provide comparable sensitivity to NP as high-$p_T$ searches.

If the NP contributions to neutral meson mixing do not have either a loop suppression or CKM suppression (or neither), then the scale sensitivity is much higher, extending to thousands of TeV.  It is indeed very easy to add NP to the SM, well outside the energy range of any current or future collider, which could still have an observable impact in flavor physics measurements.

\begin{figure*}[bt!]
\includegraphics[width=.95\columnwidth,clip,bb=15 15 550 470]{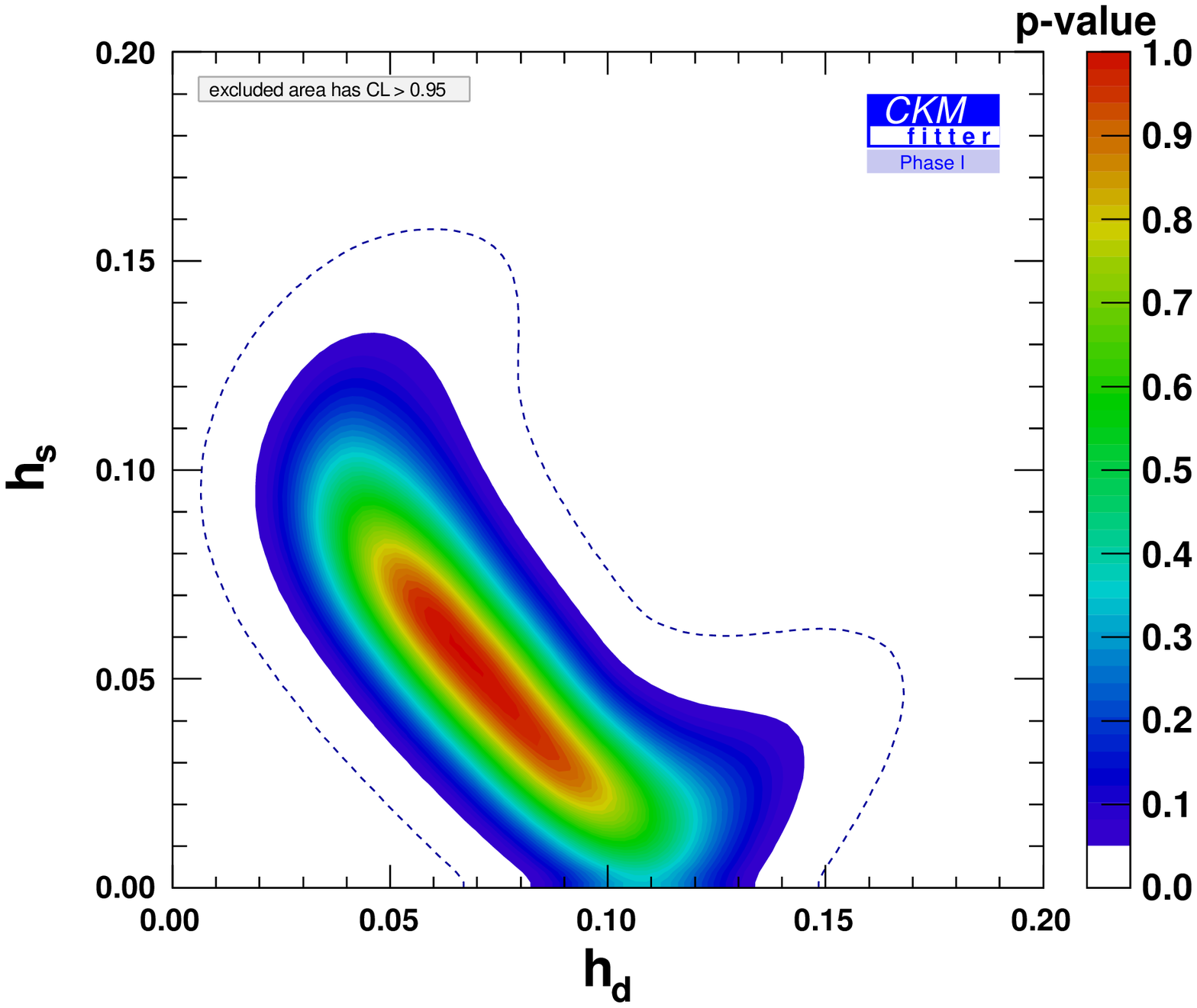}
\hfil
\includegraphics[width=.95\columnwidth,clip,bb=15 15 550 470]{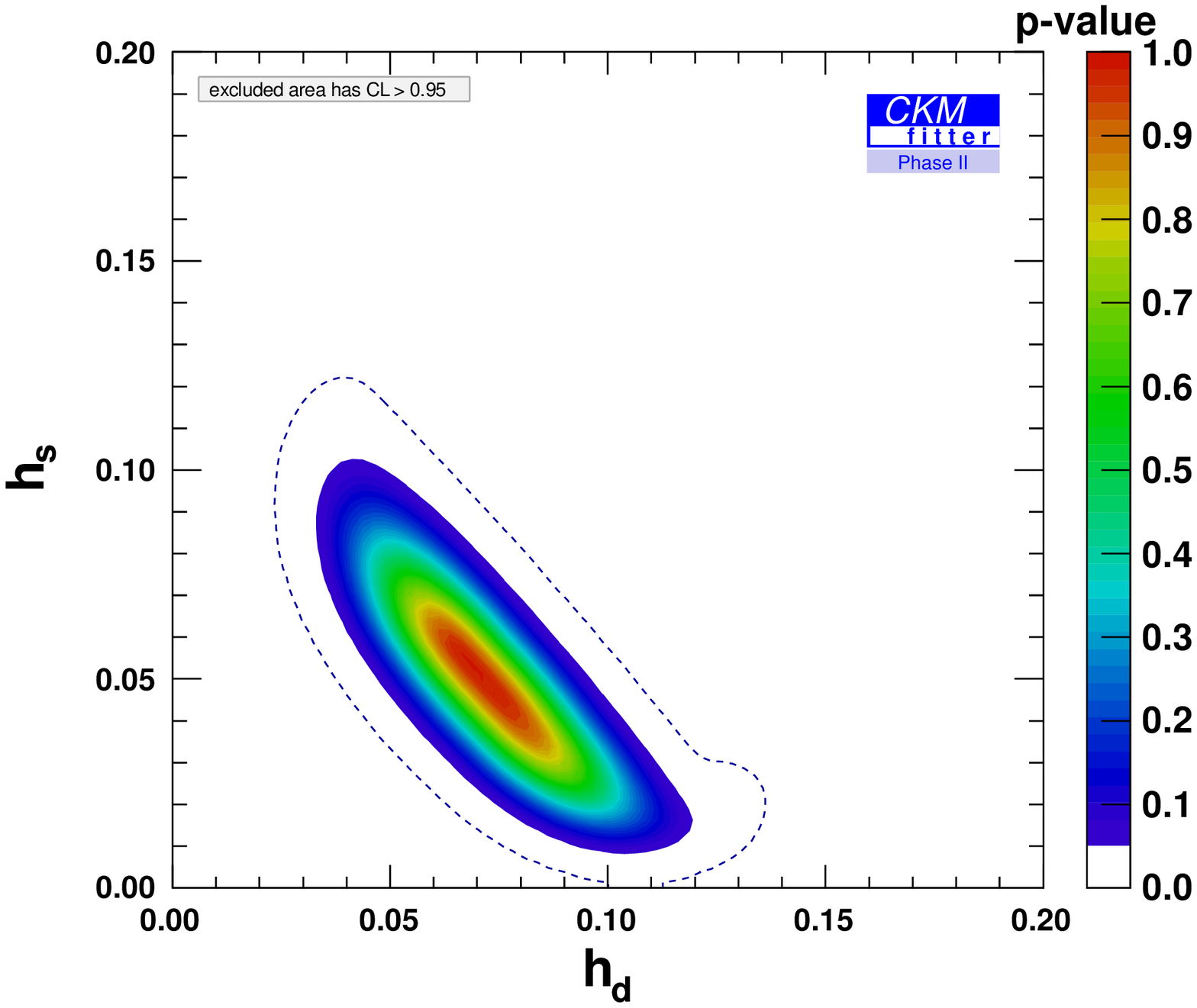}
\caption{Discovery prospects at Phase~I (left) and Phase~II (right), if the central values are as in the Summer 2019 fit in Fig.~\ref{hdhs_0}.
}
\label{hdhs_discovery}
\end{figure*}

So far in this paper we assumed that future measurements agree with the SM predictions. However, future data can not only set better bounds on NP, they may also reveal deviations from the SM.  This is illustrated in Fig.~\ref{hdhs_discovery}, where we set the CKM parameters as well as $h_{d,s}$ and $\sigma_{d,s}$ to their current best fit values (allowing for NP in $\Delta F=2$; i.e., the point indicated by the black dot in Fig.~\ref{hdhs_0}), and performed a fit assuming for all future measurements the corresponding central values, but uncertainties
as given in Table~\ref{bigtable} for Phases~I and II.  While any assumption about possible future NP signals include a high degree of arbitrariness,
Fig.~\ref{hdhs_discovery} gives an impression of the sensitivity to reveal a deviation from the SM.

%% file: discussion_inputs_bottlenecks.tex
\section{Perspectives and limitations on $|V_{cb}|$ improvement}
\label{sec:bottlenecks}

For our analysis, precise determinations of CKM parameters from tree-level
measurements is essential. This is particularly important when one reaches the Phase~II precision, where we identified the $|V_{cb}|$ precision (together with 
the  hadronic mixing parameters) as the bottleneck for this analysis. This section aims at sketching novel ways to measure the matrix element $|V_{cb}|$.

Currently there is an approximately $3\sigma$ tension in the measurements of $B\to D^{(*)}\ell\nu$ semileptonic decays, when the rate to $\tau$ leptons is
compared to the average of the $e,\, \mu$ modes~\cite{Amhis:2019ckw,Bifani:2018zmi}.  Furthermore, the inclusive $|V_{cb}|$ and $|V_{ub}|$ measurements also differ by 
more than $3\sigma$ from their exclusive counterparts~\cite{Aoki:2019cca}, when considered together. 

The tension between inclusive and exclusive
$|V_{ub}|$ and $|V_{cb}|$ determinations might also remain a cause for concern~\cite{Tanabashi:2018oca}.
If these discrepancies are not resolved and further established with higher significance in the coming years (by LHCb and Belle~II), they would also impact the
analysis of NP contributions to meson mixing.

\subsection{Opportunities for more precise \texorpdfstring{$|V_{cb}|$}{Vcb} determinations}

There are several possible ways to make progress in the future concerning $|V_{cb}|$. For the determination of $|V_{cb}|$ from exclusive semileptonic decays, separately measuring and computing the isospin difference between $B^0$ and 
$B^\pm$ decays would allow to cross check the experimental analysis against lattice QCD, whose systematic uncertainties (after including electromagnetic corrections) can reach sub-percent level. 
For the inclusive determination, similar isospin tests would be even more stringent, since isospin breaking effects 
are also suppressed by powers of $\lqcd/m_b$, and should hence be negligible, with the only outstanding issues coming from isospin-breaking induced by QED 
radiative corrections, which require matching theoretical calculations with the  specific setup of experimental analyses.

The direct determination of the CKM matrix element $|V_{cb}|$ at a high-luminosity $W$ factory (FCC-ee) has been used as an input for Phase~III.  Given the anticipated number of $W$ decays~\cite{Abada:2019lih}, the ultimate statistical precision that can be achieved is of ${\cal O}(10^{-4})$ which corresponds to about two orders of magnitude improvement with respect to the current precision. 
The key ingredient for this kind of measurements is the capability of the $c$- and $b$-jet tagging algorithms to reject the lighter quark flavours. 

Another opportunity for improvement can arise at a high-luminosity $Z$-factory, such as FCC-ee, where $B_c \to \tau \nu$ decay can be reconstructed and a measurement of the product of the $B_c$ production fraction times the branching fraction of interest  can be expected at the level of 1\%~\cite{FCC_Prell}. The challenge of the interpretation of the measurement stands in the knowledge of the $B_c$ production fraction at the $Z$ pole, where no input from $B$ factories exists. Its determination must rely on the theoretical prediction of exclusive decay branching fractions or their experimental measurement at an $e^+e^-$ collider. The $B_c^{(*)}$ pair production cross section near threshold is of the order of ${\cal O}({\rm few~fb})$~\cite{Berezhnoy:2016etd} (dominated by the $VV$ and $VP$ channels), and therefore would require the collection of ${\cal O}(10/{\rm ab})$ around $\sqrt{s}\sim 15\,{\rm GeV}$.

\subsection{If NP contributes to semileptonic \texorpdfstring{$B$}{B} decay}

The discrepancies in semileptonic $B$ decay measurements may be due to currently underestimated theoretical or experimental uncertainties or could potentially be (at least partially) explained by the
presence of BSM contributions in charged current processes. In particular, NP in the $\ell = \tau$ channel may yield violations of lepton flavor universality
(LFU), while NP in the $\ell = e,\mu$ channels can both produce LFU deviations and potentially contribute to the inclusive vs.\ exclusive tensions. The $\tau$
case  has been extensively investigated due to the fact that BSM models explaining the $R(D^{(*)})$ anomalies by modifying semi-tauonic processes are less
constrained by other measurements~\cite{Buttazzo:2017ixm,Azatov:2018knx,Murgui:2019czp,Bardhan:2019ljo,Asadi:2019xrc,Aebischer:2018iyb,Freytsis:2015qca,Dorsner:2016wpm}. 
On the other hand, BSM contributions in the $e,\mu$ channels have received less 
attention~\cite{Jung:2018lfu,Colangelo:2016ymy,Crivellin:2014zpa,Faller:2011nj,Dassinger:2007pj,Fox:2007in}. In
particular, the question of the maximum size of the NP-induced deviations in these observables achievable in viable models that respect
other experimental constraints has not been fully studied. This is relevant here, as it also corresponds to a violation of one of the assumptions of the analysis performed in this paper, namely that charged current processes are not significantly affected by BSM physics. 
Nevertheless, we now show that the $h_{d,s}-\sigma_{d,s}$ fit is still relevant for this particular scenario.

If the current anomalies in $b\to c\tau\bar\nu$ decays are attributed to BSM physics, that would imply that NP must exist at or below the TeV scale. Depending on the specific
model that is UV-completing the dimension-6 operators, ATLAS and CMS should have a good chance to directly produce and detect the particle(s) mediating
such charged current interactions. On the other hand, in some of these models, direct high-$p_T$ searches may not fully exclude BSM contributions at a level that they could still affect $|V_{cb}|$ and $|V_{ub}|$ measurements at a precision attainable in Phases~I--III.

In this case, complementary flavor physics observables can provide further insights. For example, if the NP contributions to $b\to c\ell\nu$  transitions has a different
Lorentz structure than the SM, it will manifest itself by modifying kinematic properties of the decays, e.g., the charged lepton energy or $q^2$ spectrum, the $\tau$
polarization, etc. Such NP effects may be therefore disentangled from the SM pure $V-A$ contributions, and one could in principle perform a combined fit and extract
$|V_{cb}|$ while constraining NP in semileptonic transitions. In such case the future precision to which $|V_{cb}|$ (and similarly $|V_{ub}|$) will be known, is going to be likely worse than assumed in Table~\ref{tab:CKMfitterprojinputs}. Further quantitative studies are needed to assess how much the projections performed here will be degraded and are beyond the scope of this work. 

On the other hand, if NP generates the same $V-A$ interaction as the SM (or if the contribution is smaller than what kinematic distributions can constrain), it will
bias the measurement of $|V_{cb}|$ and $|V_{ub}|$. In this case such tree-level NP effects will be seen in this analysis as nonzero contributions to
$h_{d,s}$ and $\sigma_{d,s}$, in the neutral kaon system, and in many $\Delta F=1$ processes~\cite{Fox:2007in}. These deviations from the SM induced by corrections to the charged currents will be present on top of genuine BSM FCNC contributions, which modifies the interpretation of the quantities extracted once NP is allowed and may require one ot reorganise the fit of the CKM parameters as a consequence~\cite{Descotes-Genon:2018foz}.

The fit performed here assumes the unitarity of the SM and that charged currents are only produced by SM processes. Therefore the experimental
determination of, e.g., $V_{cb}^{\rm exp} = V_{cb}^{\rm SM} + \delta V_{cb}$ is used by the fit to determine via unitarity the CKM combinations entering meson mixing, such
as $(V_{tb}V_{ts}^*)^{\rm fit}$ which are then compared with their experimental counterparts $(V_{tb}V_{ts}^*)^{\rm exp}$, with the discrepancies being attributed to
$\Delta F = 2$ NP contributions via the $h$ and $\sigma$ parameters. Similar redefinitions hold for other CKM entries determined in charged current processes, such as $|V_{ub}|$ (and to the entries in the first two generations CKM sub-matrix, although their impact is less appreciable due to the better precision to which they are known). 
This remains true as long as the tree level determination of products of $V_{ts}$, $V_{td}$, and $V_{tb}$ is not
reaching the precision attainable from FCNC processes, and therefore is inferred from unitarity. This situation will hold in the foreseeable future. So, while in the introduction we have simplified the presentation by assuming that the tree-level processes are unchanged by new physics, $h_{s,d}$ and $\sigma_{s,d}$ really parameterize generic ``tree vs.\ loop" type discrepancies.

More concretely, assuming that NP pollutes $V_{cb}$ at tree level by $\delta V_{cb}$ and similarly for $V_{ub}$, $V_{ts}$, $V_{td}$ (while neglecting contributions to $V_{tb}$, $V_{cs}$, $V_{cd}$, $V_{us}$, $V_{ud}$ for clarity), at leading order in both $m_c/m_t$ and in the size of new physics, $v^2/\Lambda^2$, we have:
\begin{eqnarray}
h_d e^{2i\sigma_d} &\simeq 2\big(V_{tb}^*\delta V_{td} + \delta V_{cb}^* V_{cd} + \delta V_{ub}^* V_{ud}\big)/(V_{tb}^* V_{td})\,,\quad  \nonumber\\
h_s e^{2i\sigma_s} &\simeq 2\big(V_{tb}^*\delta V_{ts} + \delta V_{cb}^* V_{cs} + \delta V_{ub}^* V_{us}\big)/(V_{tb}^* V_{ts})\,,\quad 
\end{eqnarray} 
which should be added to the genuine NP contributions in mixing. The full (non-linear) expressions can also be straightforwardly derived. Notice that the presence of some $\delta V_{ij}$ do not necessarily imply the presence of others for different families $i,j$, since some of these contributions may arise from operators involving right-handed quarks (below the level that can be constrained by kinematic distributions) which are unrelated by $SU(2)_L$ symmetry.

The same tree-level induced deviations from the SM predictions will appear also in all $\Delta F=1$ processes, of similar size at 
the level of the branching ratios to the contribution to $\Delta F=2$ processes.  This can be used to 
characterize the NP contributions and potentially disentangle the effects coming from measurements in charged-current processes via unitarity from genuine loop contributions. 
Furthermore, since the precision with which most of the $\Delta F=1$ decays will be able to constrain NP is 
unlikely to reach a similar level of accuracy with which $h_{s,d}$ will be constrained in Phase~III, there exists scenarios where the 
$\Delta F=2$ NP fit may still be one of the first places where NP affecting flavor changing charged currents will show up. We leave 
the identification of suitable example models to future work.

Notice also that, in the language of the SMEFT, semileptonic $B$ decays are affected by both four-fermion operators and operators 
involving a (flavor violating) quark bilinear, covariant derivatives and Higgs fields, as such operators directly modify the $Wbc$ 
vertex at order $v^2/\Lambda^2$. Therefore, the above discussion applies to the combined effect of all such operators during Phase~I and II, as the most precise determinations of $|V_{cb}|$ are at low energy. On the other hand, at Phase~III, $|V_{cb}|$
will also be well measured via $W$ decays at FCC-ee. Such measurement will not be affected by four-fermion operators and the operators whose effects cannot be disentangled in the $|V_{cb}|$ measurements will only be those modifying the $Wbc$ vertex.

%% file: summary.tex
\section{Summary and outlook}
\label{sec:sum}

The constraints on new physics in $B_d$ and $B_s$ mixings have been determined in light of recent measurements, in particular from the LHCb experiment and the $B$ factories. These results update those published in Refs.~\cite{Charles:2015gya,CKMfitterwebsite}. A good agreement with the SM is obtained, with an increased precision compared to our earlier results. As shown in Fig.~\ref{hdhs_0}, up to $\sim 20\%$ NP contributions to the mixing amplitudes are still allowed, relative to the SM contributions, and press to consider the prospects at future facilities.

The long-term experimental prospects for flavour physics involve now three proposals, with different timelines and maturity: the LHCb Upgrade~II at the high luminosity LHC, the recently initiated possible Belle~II upgrade, and finally the FCC-ee machine including a high-luminosity $Z$-factory phase to succeed the HL-LHC program at CERN.
We found that if no NP signal is seen, the
bounds on $h_d$ and $h_s$ will improve by about a factor of 3 after the first LHCb upgrade and the Belle~II completions (Phase~I), in line with the results obtained in Ref.~\cite{Charles:2013aka}, confirming the impressive progress expected from the LHCb upgrade and the Belle~II experiment. 

Though steady improvements in precision of the main observables are achieved at each of the benchmark Phases considered, they do not fully reflect into the $h_d$ and $h_s$ sensitivities. We identified these bottlenecks in precision to the determination of both $|V_{cb}|$ and the hadronic parameters of neutral-meson mixing. In relation with this question, and motivated by the tension with lepton flavour universality in $B\to D(^*)\ell\nu$, we discussed how the future facilities could improve on $|V_{cb}|$, and how it would be affected by NP in semileptonic $B$ decays.
Mixing observables have historically been a place of essential discoveries establishing the standard model, and provided crucial constraints on new physics model building. They will continue to play similar fundamental roles in the future.

\begin{acknowledgments}
We thank Y.~Grossman and A.~Poluektov for helpful conversations on $\gamma$.
We would like to thank all members of the CKMfitter group for suggestions on various aspects of this paper.
ZL was supported in part by the Office of High Energy Physics of the U.S.\ Department of Energy under contract DE-AC02-05CH11231. MP is grateful for the support provided by the Walter Burke Institute for Theoretical Physics.
LVS was supported in part by the Spanish Government and ERDF funds from the EU Commission [Grant FPA2017-84445-P] and the Generalitat Valenciana [Grant Prometeo/2017/053].

\end{acknowledgments}